\documentclass{ws-procs961x669}            
\usepackage{graphicx} 
\usepackage{booktabs} 
\usepackage{color,soul,slashed}
\usepackage{hyperref}
\usepackage[small,nohug]{diagrams}
\diagramstyle[labelstyle=\scriptstyle]

\newcommand{\CM}{{\mathbb C}}

\newcommand{\NM}{{\mathbb N}}
\newcommand{\PM}{{\mathbb P}}

\newcommand{\ZM}{{\mathbb Z}}

\newcommand{\Rr}{{\mathcal R}}

\newarrow {Congruent} 33333

\begin{document}
\title{Topological Insulators at Strong Disorder}

\author{Emil Prodan}

\address{Physics Department, Yeshiva University, New York, NY, 10016, USA.}

\begin{abstract}
Topological insulators are newly discovered materials with the defining property that any boundary cut into such crystal supports spectrum which is immune to the Anderson localization. The present paper summarizes our efforts on the rigorous characterization of these materials in the regime of weak and strong disorder. In particular, the defining property is rigorously proven under certain relevant conditions, for more than half of the classification table of topological insulators.
\end{abstract}


\bodymatter

\section{Introduction} 

A topological insulator can be defined as a homogeneous crystal with the following physical characteristics: 1) It is insulating in the bulk, that is, the diagonal components of the bulk conductivity tensor vanish in the limit of zero temperature. 2) The crystal with a boundary, however, continues to conduct electricity even in this limit. 3) These features are robust at least in the regime of small disorder. The physics community put forward a set of conjectures which were confirmed by a large body of theoretical and computational work. They say that there are only three fundamental symmetries which can stabilize topological insulating phases. The combinations of these symmetries  lead to precisely 10 classes of topological insulators and superconductors, as summarized in Table~1. \cite{AltlandPRB1997hg,SchnyderPRB2008qy,Kitaev2009hf,RyuNJP2010tq} Furthermore, it is conjectured that topological phases exists only in certain dimensions, depending on the class, and in such dimensions the phases can be labeled uniquely by either an integer number or by an integer modulo 2. When the classification is by $\ZM$ (or $2\ZM$), it is further conjectured that the integer label is provided by the Chern number of the Fermi projection in the even dimensions, and by the Chern number of the Fermi unitary operator in the odd dimensions.

\begin{table}\label{Table1}
\tbl{Classification table of strong topological insulator and superconductors. Each row represents a universality symmetry class, defined by the presence ($1$ or $\pm 1$) or absence ($0$) of the three symmetries: time-reversal (TRS), particle-hole (PHS) and chiral (CHS), and by how TRS and PHS transformations square to either $+1$ or $-1$. Each universality class is identified by a Cartan-Altland-Zirnbauer (CAZ) label. The strong topological phases are organized by their corresponding symmetry class and space dimension $d=0,\ldots,8$. These phases are in one-to-one relation with the elements of the empty, $\mathbb Z_2$, $\mathbb Z$ or $2\, \mathbb Z$ groups. The table is further divided into the complex classes A and AIII (top two rows), which are classified by the complex $K$-theory, and the real classes AI, \ldots, CI (the remaining 8 rows), which are classified by the real $K$-theory. The phases highlighted in red are the ones cover by the results reported in this paper.}
{\begin{tabular}{|c|c|c|c||c||c|c|c|c|c|c|c|c|}
\hline
$j$ & TRS & PHS & CHS & CAZ & $0,8$ & $1$ & $2$ & $3$ & $4$ & $5$ & $6$ & $7$
\\\hline\hline
$0$ & $0$ &$0$&$0$& A  & \textcolor{red}{$\ZM$} &  & \textcolor{red}{$\ZM$} &  & \textcolor{red}{$\ZM$} &  & \textcolor{red}{$\ZM$} &  
\\
$1$& $0$&$0$&$ 1$ & AIII & & \textcolor{red}{$\ZM$} &  & \textcolor{red}{$\ZM$}  &  & \textcolor{red}{$\ZM$} &  & \textcolor{red}{$\ZM$}
\\
\hline\hline
$0$ & $+1$&$0$&$0$ & AI &  \textcolor{red}{$\ZM$} & &  & & \textcolor{red}{$2 \, \ZM$} & & $\ZM_2$ & $\ZM_2$
\\
$1$ & $+1$&$+1$&$1$  & BDI & $\ZM_2$ &\textcolor{red}{$\ZM$}  & &  &  & \textcolor{red}{$2 \, \ZM$} & & $\ZM_2$
\\
$2$ & $0$ &$+1$&$0$ & D & $\ZM_2$ & ${\ZM_2}$ & \textcolor{red}{$\ZM$} &  & & & \textcolor{red}{$2\,\ZM$} &
\\
$3$ & $-1$&$+1$&$1$  & DIII &  & $\ZM_2$  &  $\ZM_2$ &  \textcolor{red}{$\ZM$} &  & & & \textcolor{red}{$2\,\ZM$}
\\
$4$ & $-1$&$0$&$0$ & AII & \textcolor{red}{$2 \, \mathbb Z$}  & &  $\ZM_2$ & $\ZM_2$ & \textcolor{red}{$\ZM$} & & &
\\
$5$ & $-1$&$-1$&$1$  & CII & & \textcolor{red}{$2 \, \ZM$} &  & $\mathbb Z_2$  & $\mathbb Z_2$ & \textcolor{red}{$\ZM$} & &
\\
$6$ & $0$ &$-1$&$0$ & C&  &  & \textcolor{red}{$2\,\ZM$} &  & $\ZM_2$ & ${\ZM_2}$ & \textcolor{red}{$\ZM$} &
\\
$7$ & $+1$&$-1$&$1$  &  CI &  & &   & \textcolor{red}{$2 \, \ZM$} &  & $\ZM_2$ & $\ZM_2$ & \textcolor{red}{$\ZM$}
\\
[0.1cm]
\hline
\end{tabular}}
\end{table}

In this paper I summarize resent results which confirm these conjectures for the topological phases highlighted in Table~1. More precisely: 1) The bulk invariants defining these phases are robust against strong disorder. As a result, to pass from one phase to another, one necessarily has to cross a localization-delocalization phase transition. 3) When a boundary is cut into a topological crystals, delocalized boundary spectrum emerges at the Fermi level. 

This work was in collaboration with Jean Bellissard and Hermann Schulz-Baldes and it was recently published as a monograph in \refcite{ProdanSpringer2016ds}. The readers interested in the proofs of the results, are directed to this reference.

\section{The Settings}

Let us start with the settings for the bulk. The disorder will be encoded in the classical dynamical system $(\Omega,\tau,\mathbb Z^d)$, where $\Omega = \Omega_0^{\ZM^d}$, $(\tau_y \omega)_x = \omega_{x+y}$ and $\Omega_0$ is a compact topological space. We assume the existence of an ergodic and invariant probability measure on $\Omega$,
$d \PM(\omega) = \prod_{x \in \ZM^d} d\PM_0(\omega_x)$, providing the averaging procedure for disorder. Above, $d \PM_0$ is a probability measure on $\Omega_0$. Throughout, $U_y$ will denote the magnetic translations: $U_x U_y = e^{\imath \pi x \wedge y}U_{x+y}$, where the skew product $x \wedge y$ encodes as usual the presence of a uniform magnetic field. The dynamics of the electrons is governed by the covariant family of finite range Hamiltonians $H= \{H_{\omega}\}_{\omega \in \Omega}$, defined over $\mathbb C^N \otimes \ell^2(\mathbb Z^d)$ and taking the generic form 
$$
H_{\omega} = \sum_{x,y \in \Rr \subset \ZM^d}  W_y(\tau_x \omega) \otimes |x \rangle \langle x  |U_y, \quad U_y H_{\omega}U_y^{-1} = H_{\tau_y \omega}, \quad |\Rr| < \infty, 
$$
where $W_y$ are continuous functions over $\Omega$. It is useful to view $W_y$ as elements of the $C^\ast$-algebra $M_{N}(\CM) \otimes C(\Omega)$ endowed with the sup norm. We will say that $H$ has a spectral gap $\Delta$ if $\sigma(H_\omega) \cap \Delta = \emptyset$ for all $\omega \in \Omega$. We will also say that $H$ has a mobility gap $\Delta$ if $\Delta$ is located in the essential spectrum of $H$ and Aizenman-Malchanov criterium applies,\cite{Aizenmann1993uf} namely, for any $s \in (0,1)$ and $\delta >0$, 
\begin{equation}\label{Eq-MABulk}
\int_{\Omega} d\PM(\omega) \, \big |\langle x,\alpha| (H_\omega-z)^{-1} |y,\beta \rangle\big |^s \leq A_s(\delta) e^{-\beta_s(\delta) |x-y|},
\end{equation}
for all $z\in \CM \setminus \sigma(h)$ with ${\rm dist}(z, \sigma(h) \setminus \Delta) \geq \delta$. Above, $A_s(\delta)$ and $\beta_s(\delta)$ are strictly positive and finite constants, which can depend parametrically on $s$ and $\delta$ but are independent of $x$ or $y$. We will say that the spectrum is delocalized if the Aizenman-Malchanov criterium cannot be established.

In the presence of a boundary, always located at $x_d =0$, the dynamics of the electrons is governed by the generic family of Hamiltonians $\widehat H_\omega = \Pi_d H_\omega \Pi_d^\ast + \widetilde H_\omega$, where $\Pi_d : \ell^2(\mathbb Z^d) \rightarrow  \ell^2(\mathbb Z^{d-1}\times \mathbb N)$ is the obvious partial isometry between these spaces, $\Pi_d H_\omega \Pi_d^\ast $ represents $H_\omega$ with the Dirichlet boundary condition and $\widetilde H_\omega$ is a generic boundary term
$$
\widetilde H_\omega \;=\;   \sum_{n,m<R} \; \sum_{y, x\in \Rr \subset \ZM^{d-1}}
  W^{y}_{nm}(\tau_{x,n} \omega) \otimes |x,n\rangle \langle x,n| U_{y,n-m}, \quad R,|\Rr| < \infty,
$$
which redefines the boundary condition. The family $\widehat H = \{\widehat H_\omega\}_{\omega \in \Omega}$ remains covariant w.r.t the lattice shifts parallel to the boundary, but the dynamical system defined by these shifts is no longer ergodic w.r.t. $d \PM$. For this reason, our results apply only for boundary disorder, encoded in an ergodic dynamical system of the type $(\Omega, \tau|_{\ZM^{d-1}},\ZM^{d-1}, d \PM_L)$, where $d\PM_L$ is the push forward measure induced by the following continuous map:
$$
p_L : \Omega_0^{\ZM^{d-1}\times I_L} \rightarrow \Omega_0^{\ZM^d}, (p_L(\omega_L))_x = \left \{
\begin{array}{l}
\omega_x \ {\rm if} \ x_d \in I_L = \{-L,\ldots,L\} \\
\omega_x =0, \ {\rm otherwise},
\end{array}
\right .
$$
where $\Omega_0^{\ZM^{d-1}\times I_L}$ is considered with the product measure $\prod_{x \in \ZM^{d-1} \times I_L} d\PM_0(\omega_x)$. Under this probability measure, the disorder configurations where any of $\omega_x$ are different from zero for $x_d>L$ occur with zero probability.

\section{Technical statements}

Let us start with the index theorems for the bulk case. It will be useful to form a matrix $\langle x | A |x \rangle$ out of the matrix elements $\langle x,\alpha |A| x,\beta \rangle$ of an operator over $\CM^N \otimes \ell^2(\ZM^d)$. We will denote by ${\rm tr}$ the trace of these matrices and will employ the Schatten norms $\|\cdot \|_{(s)}$ for these matrices. 

\begin{theorem}[Index Theorem for Bulk Projections (\refcite{BELLISSARD:1994xj,ProdanJPA2013hg})]\label{Th-IndBulkP}
Let $d$ be even and let $\{P_\omega\}_{\omega \in \Omega}$ be a family of covariant projections over $\CM^N \otimes \ell^2(\ZM^d)$ such that
$$
\sum_{x \in \ZM^d} (1+|x|)^{d+1}\int_\Omega d \mathbb P(\omega) \ \big \| \langle x \big | P_\omega |x \rangle \big \|_{(d+1)}^{d+1}  <  \infty.
$$
Consider the operator 
$$ F_{\omega,x_0} = P_\omega \left (\frac{(X+x_0) \cdot \Gamma}{|X+x_0|} \right ) P_\omega = \begin{pmatrix} 0 & G^\ast_{\omega,x_0} \\ G_{\omega,x_0} & 0 \end{pmatrix}, \quad x_0 \in (0,1)^d,
$$
where $\Gamma=(\Gamma_1, \ldots,\Gamma_d)$ is an irreducible representation of the $d$-dimensional complex Clifford algebra $\CM_d$ and $X$ is the position operator over $\ell^2(\ZM^d)$. The second equality gives the decomposition of $F_{\omega,x_0}$ w.r.t. to the natural grading of $\CM_d$. Then $G_{\omega,x_0}$ is $\PM$-almost surely a Fredholm operator on the range of $P_\omega$. Its almost sure Fredholm index is independent of $x_0 \in \ZM^d$ and $\PM$-almost surely independent of $\omega \in \Omega$, and is given by the formula
$$
{\rm Ind} \, G_{\omega,x_0} = \Lambda_d \sum_{\rho \in S_d} (-1)^\rho \int_\Omega d \PM(\omega) \ {\rm tr}\Big\langle 0 \Big| P_\omega \prod_{i=1}^d \imath \big [P_\omega,X_{\rho_i} \big] \Big |0 \Big \rangle, \quad \Lambda_d = \frac{(2\imath \pi)^{\frac{d}{2}}}{\frac{d}{2}!}.
$$
\end{theorem}

\begin{theorem}[Index Theorem for Bulk Unitaries (\refcite{Kellendonk:2002of,ProdanOddChernArxiv2014})]\label{Th-IndBulkU} 
Let $d$ be odd and let $\{U_\omega\}_{\omega \in \Omega}$ be a family of covariant unitary operators over $\CM^N \otimes \ell^2(\ZM^d)$ such that
$$
\sum_{x \in \ZM^d} (1+|x|)^{d+1}\int_\Omega d \PM(\omega) \ \big \| \langle x \big | U_\omega |x \rangle \big \|_{(d+1)}^{d+1}  <  \infty.
$$
Let $E_{x_0}$ be the spectral projection onto the positive spectrum of $(X+x_0) \cdot \Gamma$, $x_0 \in (0,1)^d$. Then, $\PM$-almost surely, the operator $F_{\omega,x_0} = E_{x_0} U_\omega E_{x_0}$ is a Fredholm operator on the range of $E_{x_0}$. Its almost sure Fredholm index is independent of $x_0$ and $\PM$-almost surely independent of $\omega \in \Omega$, and is given by the formula
$$
{\rm Ind} \, F_{\omega,x_0} = \Lambda_d \sum_{\rho \in S_d} (-1)^\rho \int_\Omega d \mathbb P(\omega) \ {\rm tr}\Big\langle 0 \Big| \prod_{i=1}^d \imath U_\omega^\ast \big [U_\omega,X_{\rho_i} \big] \Big |0 \Big \rangle, \quad \Lambda_d = \frac{\imath\,(\imath \pi)^\frac{d-1}{2}}{d!!}.
$$
\end{theorem}

In the absence of disorder and magnetic fields, the righthand sides of these index theorems are the classical even and odd Chern numbers over the $d$-torus, written in the position representation. As such, it is natural to call the above expressions the non-commutative even and odd Chern numbers.

Let us now formulate the index theorems for the boundary. Again, it will be useful to form a matrix $\langle x | \widetilde A |x \rangle$ out of the matrix elements $\langle x,x_d,\alpha |\widetilde A| x,x'_d,\beta \rangle$, for an operator $\widetilde A$ over $\CM^N \otimes \ell^2(\ZM^{d-1}\times \NM)$. Note that this time the matrices are infinite. Recall that $\|\cdot \|_{(s)}$ represents the $s$-Schatten norms  for these matrices.

\begin{theorem}[Index Theorem for Boundary Projections (\refcite{ProdanSpringer2016ds})]\label{Th-IndBoundaryP} 
Let $d$ be odd and let $\{\widetilde P_\omega\}_{\omega \in \Omega}$ be a family of covariant projections over $\CM^N \otimes \ell^2(\ZM^d \otimes \NM)$ such that
$$
\sum_{x \in \ZM^{d-1}}(1+|x|)^d \int_\Omega d \PM_L(\omega) \ \big\| \langle x | \widetilde P_\omega |x \rangle \big \|_{(d)}^d  < \infty. 
$$
Consider the operator 
$$ \widetilde F_{\omega,x_0} = \widetilde P_\omega \left (\frac{(\widetilde X - \tilde x_0) \cdot \widetilde \Gamma}{|\widetilde X - \tilde x_0|} \right ) \widetilde P_\omega = \begin{pmatrix} 0 & \widetilde G^\ast_{\omega,\tilde x_0} \\ \widetilde G_{\omega,\tilde x_0} & 0 \end{pmatrix}, \quad \tilde x_0 \in (0,1)^{d-1},
$$
where $\widetilde X$ is the position operator over $\ell^2(\ZM^{d-1})$ and $\widetilde \Gamma = (\widetilde \Gamma,\ldots,\widetilde \Gamma_{d-1})$ is an irreducible representation of the complex Clifford algebra $\CM_{d-1}$. Then, $\mathbb P_L$-almost surely, the operator $\widetilde G_{\omega,\tilde x_0}$ is a Fredholm operator on the range of $\widetilde P_\omega$. Its almost sure Fredholm index is independent of $\tilde x_0$ and $\PM_L$-almost surely independent of $\omega \in \Omega$, and is given by the formula
$$
{\rm Ind} \, \widetilde G_{\omega,\tilde x_0} =  \Lambda_{d-1} \sum_{\rho \in S_{d-1}} (-1)^\rho  \int_\Omega d \PM_L(\omega)  \ 
 {\rm tr}\Big\langle 0 \Big| \widetilde P_\omega \prod_{i=1}^{d-1} \imath \big [\widetilde P_\omega,\widetilde X_{\rho_i} \big] \Big |0 \Big \rangle.
$$
\end{theorem}

\begin{theorem}[Index Theorem for Boundary Unitaries (\refcite{ProdanSpringer2016ds})]\label{Th-IndBoundaryU} 
Let $d$ be odd and let $\{\widetilde U_\omega\}_{\omega \in \Omega}$ be a family of covariant unitaries over $\CM^N \otimes \ell^2(\ZM^d \otimes \NM)$ such that
$$
\sum_{x \in \ZM^{d-1}}(1+|x|)^d \int_\Omega d \PM_L(\omega) \ \big\| \langle x | \widetilde U_\omega |x \rangle \big \|_{(d)}^d  < \infty. 
$$
Let $\widetilde E_{x_0}$ be the spectral projection onto the positive spectrum of $(\widetilde X -\tilde x_0) \cdot \widetilde \Gamma$, $\tilde x_0 \in (0,1)^{d-1}$. Then, $\PM_L$-almost surely, the operator $\widetilde F_{\omega,\tilde x_0} = \widetilde E_{\tilde x_0} \widetilde U_\omega \widetilde E_{\tilde x_0}$ is a Fredholm operator on the range of $\widetilde E_{\tilde x_0}$. Its almost sure Fredholm index is independent of $\tilde x_0$ and $\PM_L$-almost surely independent of $\omega \in \Omega$, and is given by the formula
$$
{\rm Ind} \, \widetilde F_{\omega,\tilde x_0} =  \Lambda_{d-1} \sum_{\rho \in S_{d-1}} (-1)^\rho   \int_\Omega d \PM_L(\omega)  \ 
 {\rm tr}\Big\langle 0 \Big| \prod_{i=1}^{d-1} \imath \widetilde U_\omega \big [\widetilde U_\omega,\widetilde X_{\rho_i} \big] \Big |0 \Big \rangle.
$$
\end{theorem}
 
\section{The Bulk-Boundary Correspondence Principle}

Let us state first the results for the even dimensions. Recall that all topological phases classified by $\ZM$ (or $2 \ZM$) in even dimensions were conjectured to be uniquely defined by the top even Chern numbers of the Fermi projections $P_\omega = \chi(H_\omega \leq \mu)$, where $\mu$ represents the Fermi level.

\begin{theorem}[Bulk-Boundary Principle for Even Dimensions (\refcite{ProdanSpringer2016ds})]
\begin{enumerate}
\item If the Fermi level is located in a mobility gap, then the conditions of Theorem~\ref{Th-IndBulkP} are satisfied and as a consequence the even Chern number 
$$
{\rm Ch}_d(P_\omega) := \Lambda_d \sum_{\rho \in S_d} (-1)^\rho \int_\Omega d \PM(\omega) \ {\rm tr}\Big\langle 0 \Big| P_\omega \prod_{i=1}^d \imath \big [X_{\rho_i},P_\omega \big] \Big |0 \Big \rangle
$$ 
is quantized and $\PM$-almost surely does not fluctuate from one disorder configuration to another.
\item The invariant remains constant under continuous deformations of the functions $W_y$ in the definition of the Hamiltonians, as long as the Fermi level remains in a mobility gap. In other words, the only way ${\rm Ch}_d(P_\omega)$ can change its quantized value is through an Anderson localization-delocalization transition.
\item Let $\widehat H_\omega$ be a covariant family of half-space Hamiltonians for $H_\omega$ and assume that $H_\omega$'s have a spectral gap. Define $\widetilde U_\omega = \exp\big( 2\pi \imath f(\widehat H_\omega)\big)$, where $f$ is a smooth function such that $f= 1,0$ below/above an arbitrarily small interval around the Fermi level. Then $\widetilde U_\omega$ satisfies the conditions of Theorem~\ref{Th-IndBoundaryU} and as a consequence the odd Chern number
$$
\widetilde{\rm Ch}_{d-1}(\widetilde U_\omega) :=\Lambda_{d-1} \sum_{\rho \in S_{d-1}} (-1)^\rho  \int_\Omega d \PM_L (\omega)  \
{\rm tr} \Big\langle 0 \Big| \prod_{i=1}^{d-1} \imath \widetilde U_\omega^\ast\big [\widetilde U_\omega,\widetilde X_{\rho_i} \big] \Big |0 \Big \rangle
 $$
 is quantized and $\PM_L$-almost surely does not fluctuate from one disorder configuration to another. It is also independent of $L$.
 
\item If the spectrum is localized at the Fermi level, then the function $f$ defining $\widetilde U_\omega$ can be deformed into a step function without violating the conditions of Theorem~\ref{Th-IndBoundaryU}. In this case $\widetilde U_\omega = I$ and its boundary invariant is zero. In other words, if $\widetilde{\rm Ch}_{d-1}(\widetilde U_\omega) \neq 0$, then the boundary spectrum at the Fermi level is necessarily delocalized.

\item Furthermore, an equality between the bulk and boundary invariants holds
$$
{\rm Ch}_d(P_\omega) = \widetilde{\rm Ch}_{d-1}(\widetilde U_\omega),
$$
which is a direct consequence of Theorem~A10 of \refcite{Kellendonk:2002of}. Combining with the previous point, we now can confirm that a non-trivial bulk invariant induces delocalized boundary spectrum at the Fermi level.  
 
\end{enumerate}
\end{theorem}

We now state the results for odd dimensions. By examining Table~1, one can see that all topological phases classified by $\ZM$ (or $2\ZM$) in odd dimensions posses the chiral symmetry. This means there exists a symmetry $J$, $J^\ast =J$, $J^2 =I$, such that $J H_\omega J^{-1} = -H_\omega$. Among other things, this symmetry constraints the Fermi level to be pinned at zero, in order to satisfy charge neutrality. To avoid states pinned at zero, hence to allow for insulating states, the fiber of Hilbert space must necessarily be even. Hence, we assume that the bulk Hamiltonians are defined over $\CM^{2N} \otimes \ell^2(\ZM^d)$ and similarly for the half-space Hamiltonians. In this case, the ground state can be encoded in the unitary operator $U_\omega$ defined over $\CM^N \otimes \ell^2(\ZM^d)$ by the following implicit expression
$$
{\rm sgn}(H_\omega) = \begin{pmatrix} 0 & U^\ast_\omega \\ U_\omega & 0 \end{pmatrix},
$$
where the decomposition above is w.r.t. the grading induced by $J$.

\begin{theorem}[Bulk-Boundary Principle for Odd Dimensions]
\begin{enumerate}
\item If the Fermi level is located in a mobility gap, then the conditions of Theorem~\ref{Th-IndBulkU} are satisfied and as a consequence the odd Chern number 
$$
{\rm Ch}_d(U_\omega) := \Lambda_d \sum_{\rho \in S_d} (-1)^\rho \int_\Omega d \PM(\omega) \ {\rm tr}\Big\langle 0 \Big|  \prod_{i=1}^d \imath U_\omega^\ast\big [U_\omega,X_{\rho_i} \big] \Big |0 \Big \rangle
$$ 
is quantized and $\PM$-almost surely does not fluctuate from one disorder configuration to another.

\item The invariant remains constant under continuous deformations of the functions $W_y$ in the definition of the Hamiltonians, as long as the Fermi level remains in a mobility gap. In other words, the only way ${\rm Ch}_d(U_\omega)$ can change its quantized value is through an Anderson localization-delocalization transition.

\item Let $\widehat H_\omega$ be a covariant family of half-space Hamiltonians for $H_\omega$ and assume that $H_\omega$ have a spectral gap. Define
$$
\widetilde P_\omega = e^{-\imath \frac{\pi}{2}f(\widehat H_\omega)}
{\rm diag}(I_N,0_N)
\,e^{\imath \frac{\pi}{2}f(\widehat H_\omega)}
$$
where $f$ is a smooth function, odd under inversion and $f=\pm 1$ above/below a small interval around $\mu$. Then $\widetilde P_\omega$ satisfies the conditions of Theorem~\ref{Th-IndBoundaryP} and as a consequence the even Chern number
$$
\widetilde{\rm Ch}_{d-1}(\widetilde P_\omega) :=\Lambda_{d-1} \sum_{\rho \in S_{d-1}} (-1)^\rho  \int_\Omega d \PM_L(\omega)  \
 {\rm tr}\Big\langle 0 \Big| \widetilde P_\omega\prod_{i=1}^{d-1} \imath \big [\widetilde P_\omega,\widetilde X_{\rho_i} \big] \Big |0 \Big \rangle
 $$
 is quantized and $\PM_L$-almost surely does not fluctuate from one disorder configuration to another. It is also independent of $L$.
 
\item If the spectrum is localized at the Fermi level, then the function $f$ defining $\widetilde P_\omega$ can be deformed into a step function without violating the conditions of Theorem~\ref{Th-IndBoundaryP}. In this case $\widetilde P_\omega = {\rm diag}(I_N,0)$ and its boundary invariant is zero. In other words, if $\widetilde{\rm Ch}_{d-1}(\widetilde P_\omega) \neq 0$, then the boundary spectrum at the Fermi level is necessarily delocalized.

\item Furthermore, the following equality between the bulk and boundary invariants holds
$$
{\rm Ch}_d(U_\omega) = \widetilde{\rm Ch}_{d-1}(\widetilde P_\omega),
$$
which is a direct consequence of Theorem~A10 of \refcite{Kellendonk:2002of}. Combining with the previous point, we now can confirm that a non-trivial bulk invariant induces delocalized boundary spectrum at the Fermi level.  
 
\end{enumerate}
\end{theorem}  

\section*{Acknowledgment} This work was supported by the U.S. NSF grant DMR-1056168.

 \bibliographystyle{plain}

\begin{thebibliography}{10}

\bibitem{Aizenmann1993uf}
M.~Aizenman and S.~Molchanov.
\newblock Localization at large disorder and at extreme energies: An elementary
  derivation.
\newblock {\em Comm. Math. Phys.}, 157:245--278, 1993.

\bibitem{AltlandPRB1997hg}
A.~Altland and M.~R. Zirnbauer.
\newblock Nonstandard symmetry classes in mesoscopic normal-superconducting
  hybrid structures.
\newblock {\em Phys. Rev. B}, 55:1142--1161, 1997.

\bibitem{BELLISSARD:1994xj}
J.~Bellissard, A.~van Elst, and H~Schulz-Baldes.
\newblock The non-commutative geometry of the {Quantum Hall-Effect}.
\newblock {\em J. Math. Phys.}, 35:5373--5451, 1994.

\bibitem{Kellendonk:2002of}
J~Kellendonk, T~Richter, and H~Schulz-Baldes.
\newblock {Edge current channels and Chern numbers in the integer quantum Hall
  effect}.
\newblock {\em Rev. Math. Phys.}, 14(1):87--119, 2002.

\bibitem{Kitaev2009hf}
Alexei Kitaev.
\newblock Periodic table for topological insulators and superconductors.
\newblock In Vladimir Lebedev and Mikhail Feigel'man, editors, {\em Adv. Theor.
  Phys.: Landau Memorial Conference}, volume 1134, pages 22--30. AIP, 2009.

\bibitem{ProdanJPA2013hg}
E.~Prodan, B.~Leung, and J.~Bellissard.
\newblock {The non-commutative n-th Chern number ($n \geq 1$)}.
\newblock {\em J. Phys. A: Math. Theor.}, 46:485202, 2013.

\bibitem{ProdanOddChernArxiv2014}
E.~Prodan and H.~Schulz-Baldes.
\newblock Non-commutative odd chern numbers and topological phases of
  disordered chiral systems.
\newblock \text{http://arxiv.org/abs/1402.5002}.

\bibitem{ProdanSpringer2016ds}
E.~Prodan and H.~Schulz-Baldes.
\newblock {\em {Bulk and boundary invariants for complex topological
  insulators: From $K$-theory to physics}}.
\newblock Springer, Berlin, 2016.

\bibitem{RyuNJP2010tq}
S.~Ryu, A.~P. Schnyder, A.~Furusaki, and A.~W. Ludwig.
\newblock Topological insulators and superconductors: tenfold way and
  dimensional hierarchy.
\newblock {\em New J. Phys.}, 12:065010, 2010.

\bibitem{SchnyderPRB2008qy}
Andreas~P. Schnyder, Shinsei Ryu, Akira Furusaki, and Andreas W.~W. Ludwig.
\newblock Classification of topological insulators and superconductors in three
  spatial dimensions.
\newblock {\em Phys. Rev. B}, 78:195125, 2008.

\end{thebibliography}

\end{document}